\documentclass[aps,prd,superscriptaddress]{revtex4-1}
\usepackage[T1]{fontenc}
\usepackage[normalem]{ulem}
\usepackage{hyperref}
\usepackage{subfigure}
\usepackage{adjustbox}
\usepackage{multirow}
\usepackage{slashed}
\usepackage{pbox}
\usepackage{makecell}
\usepackage{color}
\usepackage{lineno}
\usepackage[table]{xcolor}

\usepackage{listings}
\newcommand{\ket}[1]{|#1\rangle}
\newcommand{\mmbar}{M_\mu-\overline{M}_\mu}
\newcommand{\mmu}{{M_\mu}}
\newcommand{\mmup}{{M_\mu^P}}
\newcommand{\mmuv}{{M_\mu^V}}
\newcommand{\ammu}{{\overline{M}_\mu}}

\begin{document}

\title{Muonium to antimuonium conversion:\\
Contributed paper for Snowmass 21}

\author{Ai-Yu Bai}
\affiliation{School of Physics, Sun Yat-sen University, Guangzhou 510275, China}
\author{Yu Chen}
\affiliation{School of Physics, Sun Yat-sen University, Guangzhou 510275, China}

\author{Yukai Chen}
\affiliation{Institute of High Energy Physics, Beijing 100049, China}
\author{Rui-Rui Fan}
\affiliation{Institute of High Energy Physics, Beijing 100049, China}
\author{Zhilong Hou}
\affiliation{Institute of High Energy Physics, Beijing 100049, China}
\author{Han-Tao Jing}
\affiliation{Institute of High Energy Physics, Beijing 100049, China}
\author{Hai-Bo Li}
\affiliation{Institute of High Energy Physics, Beijing 100049, China}
\author{Yang Li}
\affiliation{Institute of High Energy Physics, Beijing 100049, China}
\author{Han Miao}
\affiliation{Institute of High Energy Physics, Beijing 100049, China}
\affiliation{University of Chinese Academy of Sciences, Beijing 100049, People's Republic of China}
\author{Huaxing Peng}
\affiliation{Institute of High Energy Physics, Beijing 100049, China}
\affiliation{University of Chinese Academy of Sciences, Beijing 100049, People's Republic of China}
\author{Alexey A. Petrov (Coordindator)}
\affiliation{Department of Physics and Astronomy
Wayne State University, Detroit, Michigan 48201, USA}
\author{Ying-Peng Song}
\affiliation{Institute of High Energy Physics, Beijing 100049, China}
\author{Jian Tang (Coordinator)}
\affiliation{School of Physics, Sun Yat-sen University, Guangzhou 510275, China}
\author{Jing-Yu Tang}
\affiliation{Institute of High Energy Physics, Beijing 100049, China}
\author{Nikolaos Vassilopoulos}
\affiliation{Institute of High Energy Physics, Beijing 100049, China}
\author{Sampsa Vihonen}
\affiliation{School of Physics, Sun Yat-sen University, Guangzhou 510275, China}
\author{Chen Wu}
\affiliation{Research Center of Nuclear Physics (RCNP), Osaka University, Japan}
\author{Tian-Yu Xing}
\affiliation{Institute of High Energy Physics, Beijing 100049, China}
\author{Yu Xu}
\affiliation{School of Physics, Sun Yat-sen University, Guangzhou 510275, China}
\author{Ye Yuan}
\affiliation{Institute of High Energy Physics, Beijing 100049, China}
\author{Yao Zhang}
\affiliation{Institute of High Energy Physics, Beijing 100049, China}
\author{Guang Zhao}
\affiliation{Institute of High Energy Physics, Beijing 100049, China}
\author{Shi-Han Zhao}
\affiliation{School of Physics, Sun Yat-sen University, Guangzhou 510275, China}
\author{Luping Zhou}
\affiliation{Institute of High Energy Physics, Beijing 100049, China}

\today

\begin{abstract}
    The spontaneous muonium to antimuonium conversion is one of the interesting charged lepton flavor violation processes. It serves as a clear indication of new physics and plays an important role in constraining the parameter space beyond Standard Model. MACE is a proposed experiment to probe such a phenomenon and expected to enhance the sensitivity to the conversion probability by more than two orders of magnitude from the current best upper constraint obtained by the PSI experiment two decades ago. Recent developments in the theoretical and experimental aspects to search for such a rare process are summarized. 
\end{abstract}

\maketitle


\section{Introduction}
\label{Sec:intro}
Neutrino oscillation is a neutral lepton flavor violation process with profound implications on particle physics. On one hand, it points out that neutrinos have mass and it is therfeore the first direct evidence of physics beyond standard model (BSM). The existence of neutral lepton flavor violation also leads to ask whether there can be also charged lepton flavor violation (cLFV). On the other hand, the origin of neutrino masses remains one of the unsolved mysteries in particle physics. The traditional way to explain the neutrino masses is the seesaw mechanism, which is often predicted together with cLFV effects~\cite{Ilakovac:1994kj}. For example, in the type-II seesaw model a Higgs triplet is introduced under a $SU(2)$ symmetry. After the spontaneous symmetry breaking, the massive Higgs boson can contribute to cLFV processes. Therefore, searching for cLFV is an important quest in BSM physics. It is well motivated to push forward the experimental efforts to look for BSM physics by cLFV processes.

\begin{figure}[!b]
\centering
\includegraphics[width=0.4\textwidth]{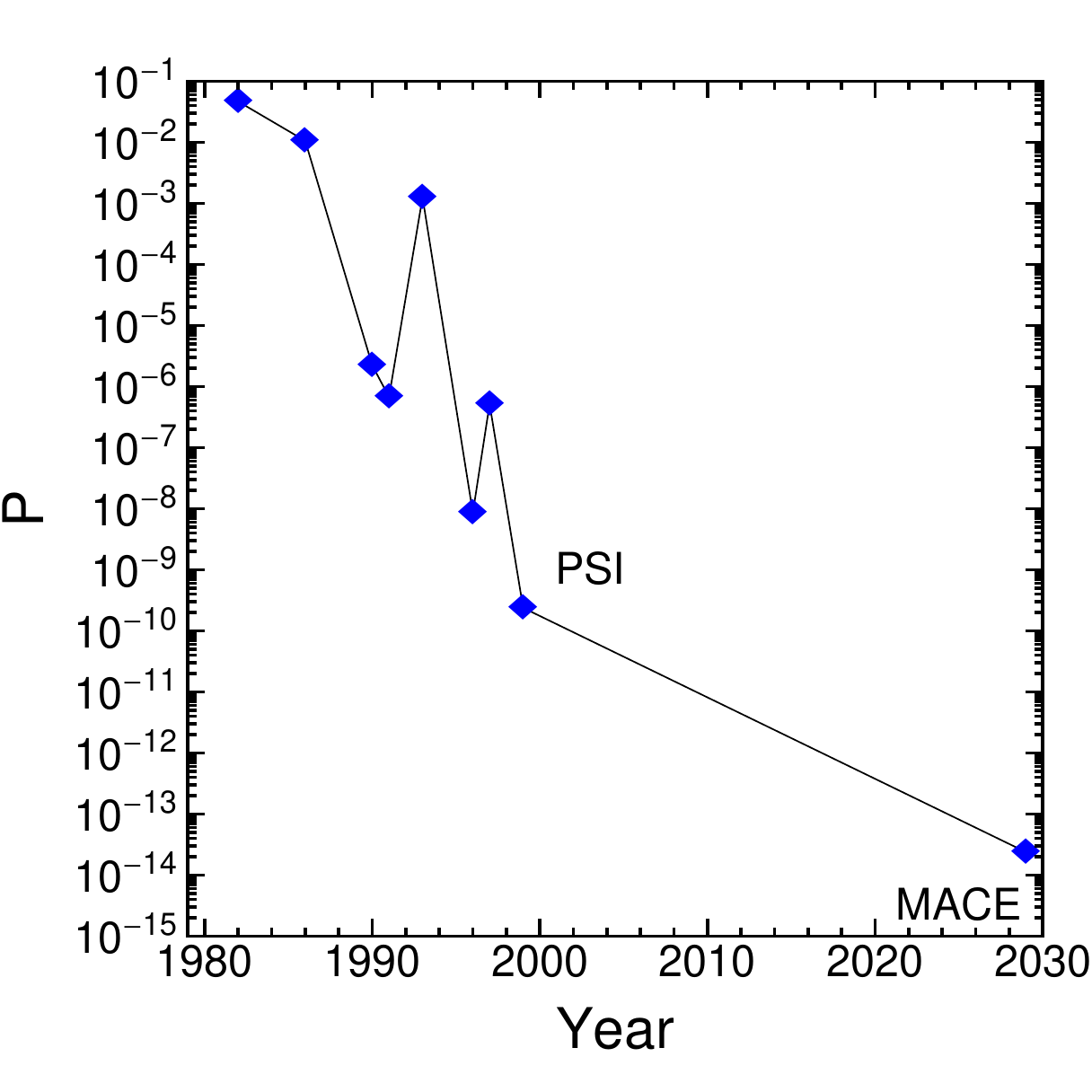}
\caption{The history of searching for the spontaneous muonium to antimuonium conversion, including the expected sensitivity of the MACE experiment.}
\label{fig:pic1}
\end{figure}

There are many experimental efforts to search for cLFV. Such cLFV experiments as \textbf{COMET}~\cite{Adamov:2018vin} in Japan and \textbf{Mu2e}~\cite{Bartoszek:2014mya} in the USA are currently under construction to search for the coherent muon to electron conversion $\mu^-N\to e^- N$. The accelerator muon beam experiments at PSI are also searching cLFV processes, with \textbf{Mu3e}~\cite{Berger:2014vba} looking for $\mu^+\to e^+ e^- e^-$ and \textbf{MEG-II}~\cite{Baldini:2018nnn} $\mu^+ \to e^+ \gamma$, respectively. Another interesting method to probe BSM physics via cLFV is to take the muonium atom and see whether there is a spontaneous conversion from muonium to antimuonium. The original idea was first brought by Pontecorvo ~\cite{Pontecorvo:1957cp}. In an effective field theory, the study of muonium to antimuonium conversion prospects can be found~\cite{Conlin:2020veq}. A more recent study presents a complete list of models to induce such a conversion process ~\cite{Fukuyama:2021iyw}. In a type-II related hybrid seesaw model, the double-charged Higgs particle predicts the muonium to antimuonium conversion even at the tree level~\cite{Han:2021nod}. The history of searching for the muonium to antimuonium process is illustrated in Fig.~\ref{fig:pic1}. The latest measurement on the muonium to antimuonium conversion probability was obtained at P $\lesssim8.3\times10^{-11}$ by 90\% confidence level in a PSI experiment in 1999. This result has not been challenged in any experiment within the past two decades.

The advent of new intense and slow muon sources and the significant advances in modern particle detection technologies will lead to new possibilities in the design of new lepton flavor violation experiments. Regarding the muonium to antimuonium conversion, we intend to improve the present bound by more than two orders of magnitude in the proposed Muonium-to-Antimuonium Conversion Experiment (MACE). The conceptual design study focuses on the following aspects: the high-efficency muonium formation, an optimized design of magnetic spectrometer and requirements in physics performance along with a development of the proper accelerator muon beam, the corresponding tracking algorithm to discriminate signals and backgrounds and a refined figure of merit to present physics sensitivity. This white paper summarizes the theoretical and experimental aspects of MACE.

\section{Theoretical aspects}
\label{Sec:th}

The main decay channel for the muonium is determined by the weak decay of the 
muon, $\mmu \to e^+e^- \bar \nu_\mu \nu_e$, so the average lifetime of a muonium state $\tau_\mmu$ is 
expected to be the same as that of the muon, 
\begin{equation}\label{MuoniumWidth}
1/\tau_\mmu = \Gamma\left(\mmu \to e^+e^- \nu_e \bar\nu_\mu \right) \approx
\Gamma\left(\mu \to e^+ \nu_e \bar\nu_\mu \right) = \frac{G_F^2 m_\mu^5}{192 \pi^3}
=1/\tau_\mu , 
\end{equation}
with $\tau_\mu = (2.1969811 \pm 0.0000022)\times 10^{-6}$ s \cite{Tanabashi:2018oca},
apart from the tiny effect due to time dilation \cite{Czarnecki:1999yj}. Note that Equation (\ref{MuoniumWidth}) represents
the leading-order result. The results including subleading corrections are available \cite{Czarnecki:1999yj}.

Like a hydrogen atom, muonium could be formed in two spin configurations. A spin-one triplet state $\ket{\mmuv}$ is
called {\it ortho-muonium}, 
while a spin-zero singlet state $\ket{\mmup}$ is called {\it para-muonium}. In what follows, we 
will drop the superscript and employ the notation $\ket{\mmu}$ if the spin of the muonium state is not important for the 
discussion.

Since $\Delta L=2$ interaction can change the muonium state into the anti-muonium one, the possibility to study 
muonium--anti-muonium oscillations arises. Theoretical analyses of conversion probability for muonium into 
antimuonium have been performed, both in particular new physics
models \cite{Pontecorvo:1957cp,Feinberg:1961zza,ClarkLove:2004,CDKK:2005,Li:2019xvv,Endo:2020mev}, and 
using the framework of effective theory \cite{Conlin:2020veq}, where all possible BSM models are encoded 
in a few Wilson coefficients of effective operators. Observation of muonium converting into anti-muonium provides clean 
probes of new physics in the leptonic sector \cite{Bernstein:2013hba,Willmann:1998gd}. 


In order to determine experimental observables related to $\mmbar$ oscillations, we recall that 
the treatment of the two-level system that represents muonium and antimuonium is similar to that 
of meson-antimeson oscillations \cite{Petrov:2021idw,Donoghue,Nierste}. There are, however, several
important differences. First, both ortho- and para-muonium can oscillate. Second, the SM oscillation
probability is tiny, as it is related to a function of neutrino masses, so any experimental indication of oscillation
would represent a sign of new physics. 
 
In the presence of the interactions coupling $\mmu$ and $\ammu$, the time development of a muonium 
and anti-muonium states would be coupled, so it would be appropriate to consider their combined evolution, 
\begin{equation}
|\psi(t)\rangle = 
\left( {\begin{array}{c}
 a(t) \\
 b(t) \\
 \end{array} } \right) =
 a(t) |\mmu\rangle + b(t) |\ammu\rangle.
\end{equation}

The time evolution of $|\psi(t)\rangle$ evolution is governed by a Schr\"odinger-like equation, 
\begin{equation}\label{two state time evolution}
i\frac{d}{dt}
\left(
\begin{array}{c}
\ket{M_\mu(t)} \\ \ket{\overline{M}_\mu(t)}
\end{array}
\right)
=
\left(m-i\frac{\Gamma}{2}\right)
\left(
\begin{array}{c}
\ket{M_\mu(t)} \\ \ket{\overline{M}_\mu(t)}
\end{array}
\right).
\end{equation}
where $\left(m-i\frac{\Gamma}{2}\right)_{ik}$ is a $2\times 2$ Hamiltonian (mass matrix) with non-zero off-diagonal
terms originating from the $\Delta L=2$ interactions. CPT-invariance dictates that the masses and widths of 
the muonium and anti-muonium are the same, so $m_{11}= m_{22}$, $\Gamma_{11}=\Gamma_{22}$.
In what follows, we assume CP-invariance of the $\Delta L_\mu = 2$ interaction. {A more general formalism without this assumption follows the same steps as that for the $B\bar B$ or $K\bar K$ mixing \cite{Petrov:2021idw,Donoghue}.}.
Then,
\begin{eqnarray}\label{off_diagonal_elements}
m_{12}=m^{*}_{21}, \qquad \Gamma_{12}=\Gamma^{*}_{21}.
\end{eqnarray}

The off-diagonal matrix elements in Equation (\ref{off_diagonal_elements}) can be related to the matrix elements of 
the effective operators introduced in Section \ref{Sec:intro}, as discussed in \cite{Petrov:2021idw,Donoghue},
\begin{equation}\label{OffDiagonal}
\left(m-\frac{i}{2} \Gamma\right)_{12}=\frac{1}{2 M_{M}}\left\langle\ammu\left|{\cal H}_{\rm eff} \right| \mmu\right\rangle
+\frac{1}{2 M_{M}} \sum_{n} \frac{\left\langle\ammu
\left|{\cal H}_{\rm eff} \right| n\right\rangle\left\langle n\left|{\cal H}_{\rm eff} \right| \mmu\right\rangle}{M_{M}-E_{n}+i \epsilon}.
\end{equation}

To find the propagating states, the mass matrix needs to be diagonalized. The basis in which the 
mass matrix is diagonal is represented by the mass eigenstates $ |\mmu_{1,2} \rangle$, which are 
related to the flavor eigenstates $\mmu$ and $\ammu$ as
\begin{equation}
 |\mmu_{1,2} \rangle = \frac{1}{\sqrt{2}} \left[ |\mmu \rangle \mp |\ammu \rangle
 \right] ,
\end{equation}
where we employed a convention where $CP |\mmu_\pm \rangle = \mp |\mmu_\pm \rangle$.
The mass and the width differences of the mass eigenstates are 
\begin{eqnarray}\label{def of mass and width difference}
\Delta m \equiv M_{1}-M_{2}, \qquad \Delta \Gamma \equiv \Gamma_{2}-\Gamma_{1}.
\end{eqnarray} 

Here, $M_i$ ($\Gamma_i$) are the masses (widths) of the physical mass eigenstates $ |\mmu_{1,2} \rangle$. 

It is interesting to see how the Equation (\ref{OffDiagonal}) defines the mass and the lifetime differences.
Since the first term in Equation (\ref{OffDiagonal}) is defined by a local operator, its matrix element 
does not develop an absorptive part, so it contributes to $m_{12}$, i.e., the mass difference. 
The second term contains bi-local contributions connected by physical intermediate states. This term 
has both real and imaginary parts and thus contributes to both $m_{12}$ and $\Gamma_{12}$.

It is often convenient to introduce dimensionless quantities,
\begin{equation}\label{XandY}
x = \frac{\Delta m}{\Gamma}, \qquad y = \frac{\Delta \Gamma}{2\Gamma},
\end{equation}
where the average lifetime $\Gamma=(\Gamma_1+\Gamma_2)/2$, and $M_M = \left(M_1+M_2\right)/2$ 
is the muonium mass. Noting that $\Gamma$ is defined by the standard model decay rate of the muon, 
and $x$ and $y$ are driven by the lepton-flavor violating interactions, we should expect that both $x,y \ll 1$.

The time evolution of flavor eigenstates follows from Equation (\ref{two state time evolution}) \cite{Donoghue,Nierste,Conlin:2020veq},
\begin{eqnarray}
\ket{M(t)} &=& g_+(t) \ket{\mmu} + g_-(t) \ket{\ammu},
\nonumber \\
\ket{\overline{M}(t)} &=& g_-(t) \ket{\mmu} + g_+(t) \ket{\ammu},
\end{eqnarray}
where the coefficients $g_\pm(t)$ are defined as
\begin{equation}\label{TimeDep}
g_\pm(t) = \frac{1}{2} e^{-\Gamma_1 t/2} e^{-iM_1t} \left[1 \pm e^{\Delta\Gamma t/2} e^{i \Delta m t} \right].
\end{equation}

As $x,y \ll 1$ we can expand Equation (\ref{TimeDep}) in power series in $x$ and $y$ to obtain
\begin{eqnarray}
g_+(t) &=& e^{-\Gamma_1 t/2} e^{-iM_1t} \left[1 + \frac{1}{8} \left(y-ix\right)^2 \left(\Gamma t\right)^2 \right],
\nonumber \\
g_-(t) &=& \frac{1}{2} e^{-\Gamma_1 t/2} e^{-iM_1t} \left(y-ix\right) \left(\Gamma t\right).
\end{eqnarray}

The most natural way to detect $\mmbar$ oscillations experimentally is by producing $\mmu$ state and looking for the decay 
products of the CP-conjugated state $\ammu$. Denoting an amplitude for the $\mmu$ decay into a final state 
$f$ as $A_f = \langle f|{\cal H} |\mmu\rangle$ and an amplitude for its decay into a CP-conjugated final state 
$\overline{f}$ as $A_{\bar f} = \langle \overline{f}|{\cal H} |\mmu\rangle$,
we can write the time-dependent decay rate of $\mmu$ into the $\overline{f}$,
\begin{equation}
\Gamma(\mmu \to \overline{f})(t) = \frac{1}{2} N_f \left|A_f\right|^2 e^{-\Gamma t} \left(\Gamma t\right)^2 R_M(x,y),
\end{equation}
where $N_f$ is a phase-space factor and we defined the oscillation rate $R_M(x,y)$ as
\begin{equation}
R_M(x,y) = \frac{1}{2} \left(x^2+y^2\right).
\end{equation}

Integrating over time and normalizing to $\Gamma(\mmu \to f)$ we get the probability of $\mmu$ decaying as $\ammu$ at some time $t > 0$,
\begin{equation}\label{Prob_osc}
P(\mmu \rightarrow \ammu) = \frac{\Gamma(\mmu \to \overline{f})}{\Gamma(\mmu \to f)} = R_M(x,y). 
\end{equation}

The equation Equation (\ref{Prob_osc}) \cite{Conlin:2020veq} generalizes oscillation probability found in the 
papers \cite{Feinberg:1961zza,CDKK:2005} by allowing for a non-zero lifetime difference in $\mmbar$ oscillations.

\section{Experimental aspects}
\label{Sec:ex}

\subsection{Potential accelerator muon beamlines}
\label{Subsec:beam}
\begin{figure}[!b]
\centering
\includegraphics[width=0.4\textwidth]{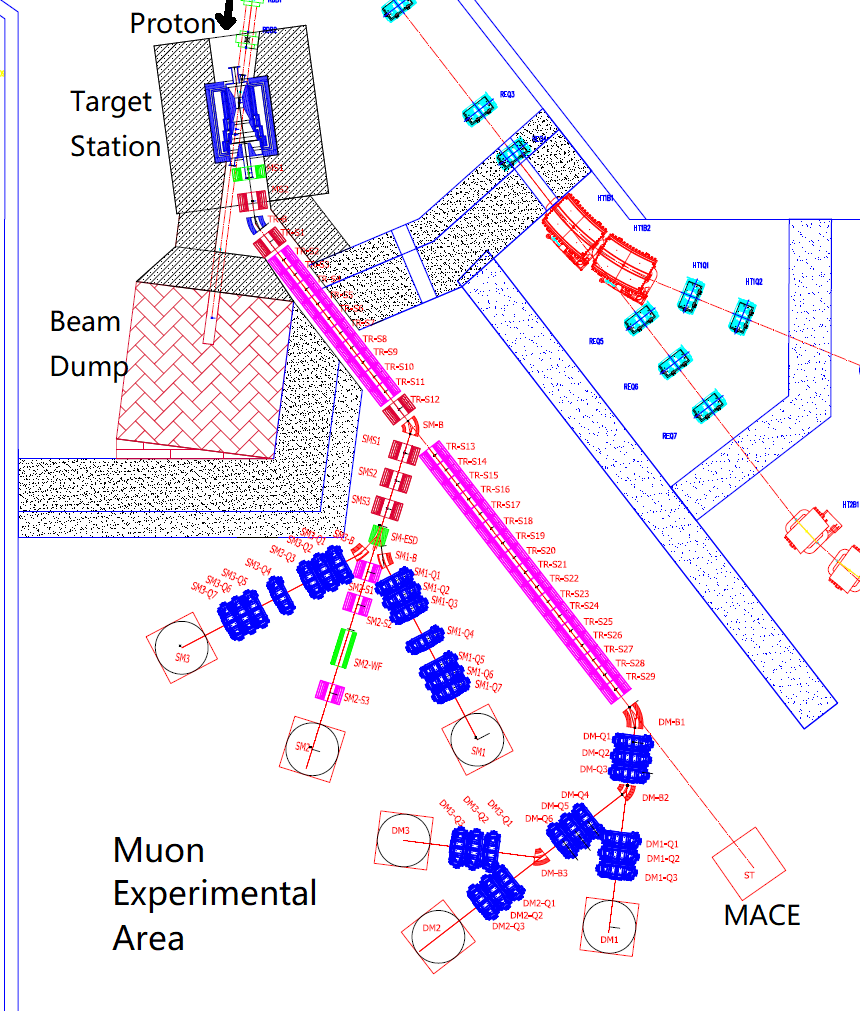}
\caption{The conceptual design of the muon beamline for the MACE experiment.}
\label{fig:EMuS}
\end{figure}

A 100 kW pulsed proton accelerator with the beam energy of 1.6 GeV and the repetition frequency of 25 Hz has been running at China Spallation Neutron Source (CSNS) in the YGA Bay area since 2018. An upgrade of the beam power toward 500 kW has been approved with a working package specifically reserved for the construction of an experimental muon source (EMuS)~\cite{qubs2040023}. Meanwhile, China Initiative Accelerator Driven sub-critical System (CiADS) will offer  continous proton beam with the power at the Megawatt level. Both accelerator facilities are willing to make use of the proton beam to offer the muon beamline. With the EMuS baseline scheme, a proton beam of 25 kW in beam power and 2.5 Hz in repetition rate will bombard a graphite target of conical shape that is located in high-field superconducting solenoids for producing different types of high-intensity muon beams. As for MACE experiment, the surface muons as high as $10^8~\mu^+$/s will be transported to the experimental area through the muon beam transportation line also based on the superconducting solenoids whose conceptual design is shown in FIG.~\ref{fig:EMuS}. We expect the beam spread smaller than 5\% in order to meet requirements of new physics searches. we also keep eyes on the global accelerator muon beamlines as is given in the TABLE~\ref{tab:muon_source}for potential chance. 

\begin{table}[!t]
\centering
\begin{tabular}{|c|c|c|c|c|c|c|c||}
\hline\hline
 & Proton driver [MW] & Intensity [$\times10^6$/s] & Polarization[\%] & Spread [\%] \\
\hline
 PSI & 1.30 & 420 & 90 & 10 \\
\hline
ISIS & 0.16 & 1.5 & 95 & $\leq15$\\
\hline
RIKEN/RAL & 0.16 & 0.8 & 95 & $\leq15$\\
\hline
JPARC & 1.00 & 100 & 95 & 15 \\
\hline
TRIUMF & 0.075 & 1.4 & 90 & 7\\
\hline
EMuS & 0.025 & 83 & 50 & 10\\
\hline\hline
\end{tabular}
\caption{\label{tab:muon_source}The potential accelerator muon sources around the world, including the proposed EMuS in China Spallation Neutron Source.}
\end{table}

\subsection{Muonium production and diffusion}
\label{Subsec:muonium prod}

\begin{figure}[!t]
    \centering
    \includegraphics[width=15cm]{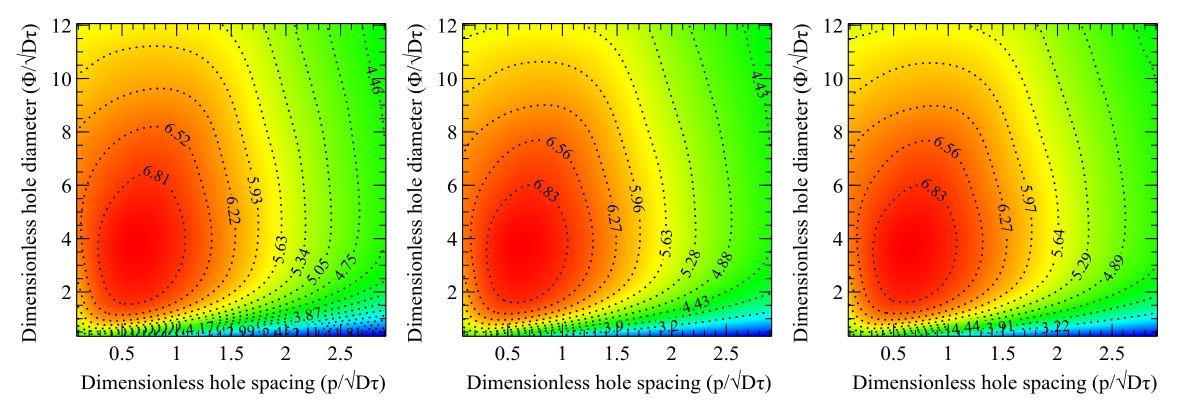}
    \includegraphics[width=15cm]{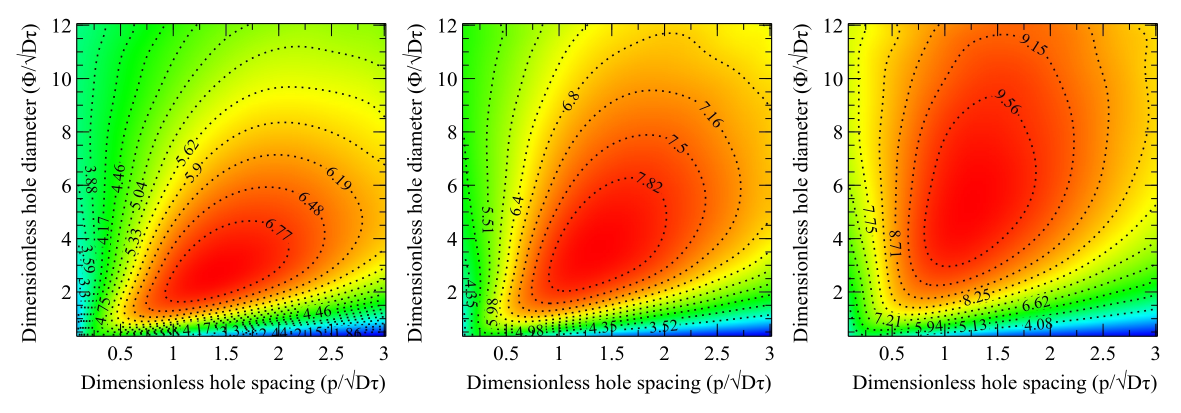}
    \caption{The relative muonium yield of laser-ablated silica aerogel target based on the Monte Carlo simulation. The yield value is relative to that of the flat target (without holes or grooves at downstream side). The left, middle, and right graphs correspond to the hole depths of 2.5~mm, 5~mm, and 7.5~mm, respectively. The top three panels correspond to the case where the beam momentum spreading is small, and the simulation assumes that muoniums are produced within a few millimeters near the surface of the target, and thus tends to diffuse out. The bottom three panels correspond to the situation where the beam momentum spreading is very large, and it is assumed that muoniums are uniformly produced in the target. }
    \label{fig:muoniumYield}
\end{figure}

In order to enhance the sensitivity of the spontaneous muonium transition event in MACE, an essential aspect is to increase the muonium yield in vacuum. As the most common approach of producing muonium in the experiment, a surface muon beam with momentum at around 25 MeV hits the target, and muons were injected into the material and were quickly slowed down to the room temperature. During the process, some of muons could spontaneously capture an electron from the material atom before the eventually decay, and form muoniums. They are excepted to emit to the vacuum as much as possible, since one of the most crucial design of MACE is to detect the very low energy atomic shell electron, which can only be detected when the muonium decays outside the material. Thus, there are two main aspect on the benefit for the vacuum yield of muoniums: the first aspect is to select the material of the target in order to obtain higher electron capture efficiency, or the absolute muonium yield, and the second aspect is to push the muonium emission rate to the maximum. Previous studies have made efforts on the latter. There were some attempts on the target design like tungsten foil target, and silica power target used in PSI experiment~\cite{Willmann:1998gd}, which obtained the electron capture efficiency of 61\% and vacuum emission rate of about 3\%. The first attempt on the silica aerogel showed its potential of higher vacuum muonium yield~\cite{Schwarz.1}. A recent muonium emission experiment by TRIUMF and J-PARC compared flat (without holes or grooves at downstream side) with laser-ablated (with holes or grooves at downstream side) aerogel target~\cite{Beare:2020gzr}. They found that the muonium emission rate of flat aerogel targets was 4 times higher than that of silica powder target, and the yield of aerogel target with holes has up to 36 times higher than that of silica powder targets. The study demonstrates that the aerogel target with specific structures like laser-ablated holes can lead to significantly higher vacuum muonium yields than silica powder target or untreated aerogel target. Base on the previous works, we carried out further research on this topic. A Monte Carlo simulation of muonium diffusion based on the random walk method is applied. As is given in FIG.~\ref{fig:muoniumYield}, we show the relative muonium yield of laser-ablated silica aerogel target in different configurations. The top three panels correspond to the case where the beam momentum spreading is small, and the simulation assumes that muoniums are produced within a few millimeters near the surface of the target, and thus tends to diffuse out. In this case, the depth of the hole has little effect on the yield, and the optimal yield can be 7.1 times higher than that of the flat target, which means that 23.3\% of muoniums can diffuse into the vacuum. The bottom three panels correspond to the situation where the beam momentum spreading is very large, and it is assumed that muoniums are uniformly produced in the target. In this case, the deeper the hole, the higher the yield. The optimal yield can be 9.9 times higher than that of the flat target, which means that 2.04\% of the muonium can diffuse into the vacuum. The result shows the very promising improvement compared with the existing technologies. Nevertheless, there is still room for further improvement on high-efficiency muonium production and diffusion.

\subsection{Magnetic spectrometer and triple coincidence system}
\label{Subsec:spectrometer}

\subsubsection{Triple coincidence system}

After diffusing from the target surface, the muonium might spontaneously convert to anti-muonium, then decays into a positron with atomic-shell energy ($\sim$13.5~eV), a electron with Michel energy (52.8~MeV max), and two invisible neutrinos. The experiment goal is to identify the anti-muonium by detecting its decay products via multiple coincidence method, and the signal event is identified when the Michel electron signal become coincident with the atomic-shell positron signal. On one hand, to detect the Michel electron, a magnetic spectrometer will measure the muonium decay vertex and particle momentum. On the other hand, a composite detector system consists of a microchannel plate (MCP) and electromagnetic calorimeter would be introduced to detect the atomic-shell positron. The atomic-shell positron will be accelerated to higher energy around a few keV, and be guided to MCP through a transportation line. The transport line ensures that the position of the positron in the vertical plane is conserved. Then the positron will hit the MCP and its position will be recorded. The positron is then expected to annihilate into two photons with energy of 0.511~MeV, which can be detected by the calorimeter simultaneously. By the triple coincidence system consists of magnetic spectrometer, MCP and calorimeter, the muonium to anti-muonium conversion event can be identified.

\begin{figure}
    \centering
    \includegraphics[width=17cm]{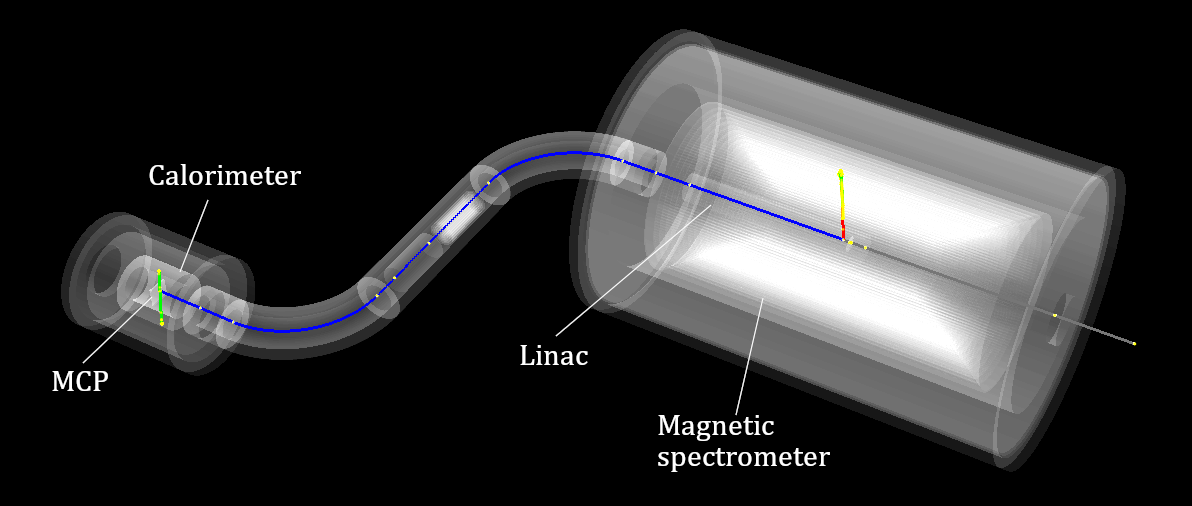}
    \caption{A Geant4 simulated signal event. The incident gray track represents a surface muon hits on the target. The Michel electron is represented by the red track, which is detected by the drift chamber. The atomic-shell positron, which is then accelerated and transported to the MCP, annihilates into two photons. The photons are detected by the electromagnetic calorimeter. }
    \label{fig:MACEsetup}
\end{figure}

\subsubsection{Magnetic spectrometer}

The spectrometer system is currently designed as a drift chamber placed inside a longitudinal magnetic field. To improve the sensitivity, discriminate muoniums and anti-muoniums, and benefit for the distinction between signal events and background events, the drift chamber should have high detection efficiency and covers a reasonably large solid angle, high spatial resolution and low error rate of charge discrimination, and high momentum resolution, respectively. These are the criteria for designing the drift chamber for MACE.

\begin{figure}
    \centering
    \includegraphics[width=12cm]{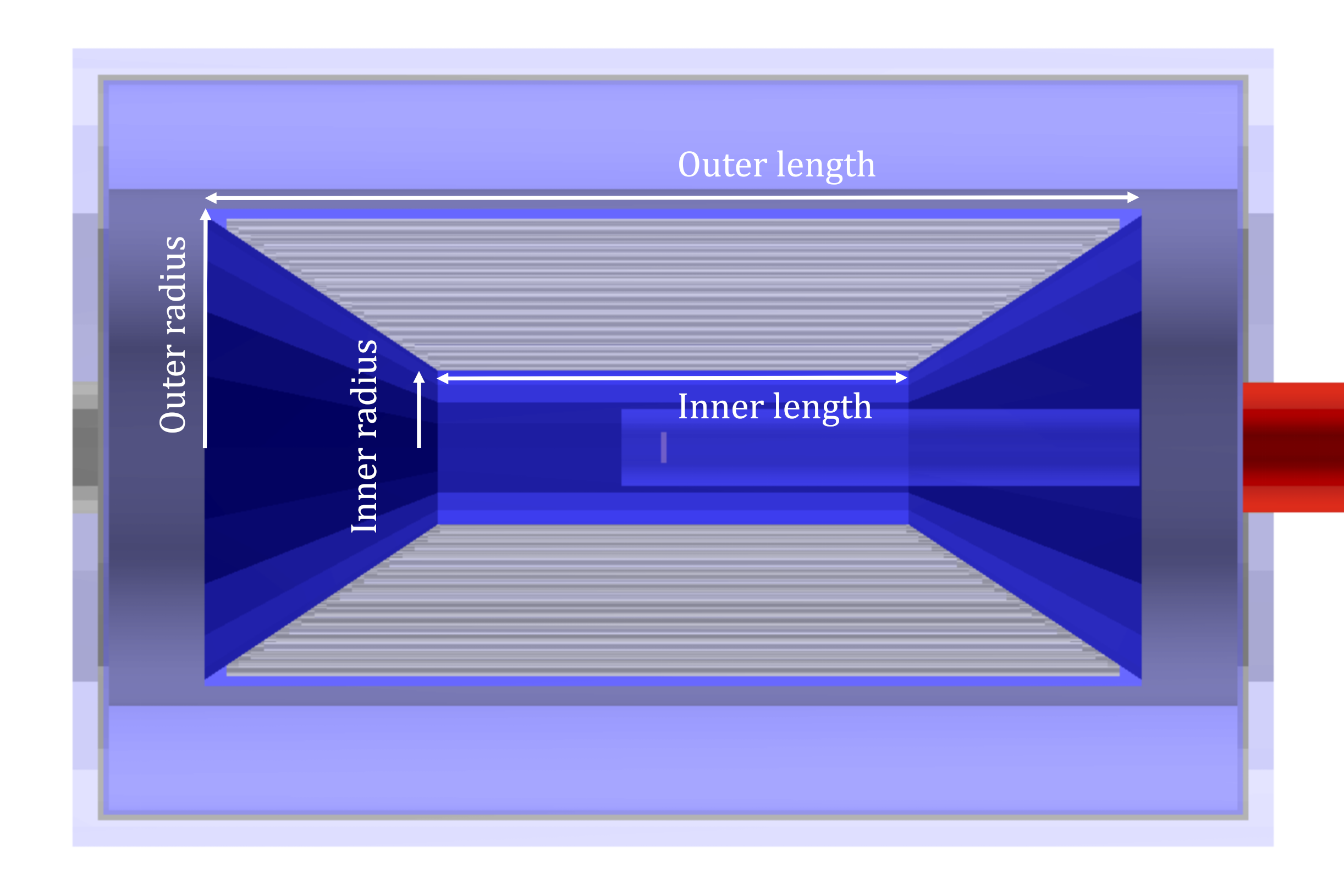}
    \caption{The clipping view of the spectrometer without an implementation of the hodoscope to facilitate the tracking and the cosmic muon veto system yet.}
    \label{fig:clipppingCDC}
\end{figure}

The performance of drift chamber has been studied by Geant4 simulation. Events are generated as true experimental setup, that is, surface muons hit the target and drift chamber detects decay positrons and electrons. The muonium to anti-muonium conversion probability is raised to 50\% to investigate the performance for both electrons and positrons. The simulation just focuses on the overall performance of the detector currently by simply recording the drift distance and longitude hit position. The detailed response is currently not considered, and wires are arranged parallel to the longitudinal axis. The drift chamber is modeled as an rotated trapezoid (see FIG.~\ref{fig:clipppingCDC}), inside which cells are constructed layer by layer, staggered by half the cell width if possible. Each cell is surrounded by a near-squared field wire grid, with the average width of 16.2~mm. One cell has 3 field wires and 1 sense wire, wire diameters used in the simulation are 110~$\mu$m and 25~$\mu$m, respectively. The drift chamber works in an longitudinal magnetic field of 0.1T. For now, drift chamber of different length and radius are investigated, with identical solid angle coverage of 95\% (inner), 89\% (outer). The length is decided by the radius and solid angle coverage. Parameters of all tested drift chamber setups are listed in TABLE~\ref{table:CDC parameters} except for identical parameters.

\begin{table}[htbp]
    \begin{center}
        \caption{Summary of investigated drift chamber setups. Setup 8 is adopted as the optimal design. See FIG.~\ref{fig:cdcPerformace} for details.}
        \label{table:CDC parameters}
        \begin{tabular}{c|cccc|ccc}
            \hline
                  & Inner & Outer & Inner & Outer & Number & Number \\
            Setup & radius (cm) & radius (cm) & length (cm) & length (cm) & of layers & of cells \\
            \hline
            1     & 10     & 30     & 60     & 120    & 12     & 952      \\
            2     & 10     & 35     & 60     & 140    & 15     & 1324      \\
            3     & 15     & 35     & 90     & 140    & 12     & 1188      \\
            4     & 10     & 40     & 60     & 160    & 18     & 1768      \\
            5     & 15     & 40     & 90     & 160    & 15     & 1596      \\
            6     & 20     & 40     & 120    & 160    & 12     & 1412   \\
            7     & 10     & 45     & 60     & 180    & 21     & 2248      \\
            \textbf{8}&\textbf{15}&\textbf{45}&\textbf{90}&\textbf{180}&\textbf{18}&\textbf{2112}      \\
            9     & 20     & 45     & 120    & 180    & 15     & 1928      \\
            10    & 25     & 45     & 150    & 180    & 12     & 1676      \\
            11    & 10     & 50     & 60     & 200    & 24     & 2808      \\
            12    & 15     & 50     & 90     & 200    & 21     & 2628      \\
            13    & 20     & 50     & 120    & 200    & 18     & 2444      \\
            14    & 25     & 50     & 150    & 200    & 15     & 2204      \\
            15    & 30     & 50     & 185    & 200    & 12     & 1876      \\
            \hline
        \end{tabular}
    \end{center}
\end{table}

\begin{figure}[!t]
    \centering
    \includegraphics[width=15cm]{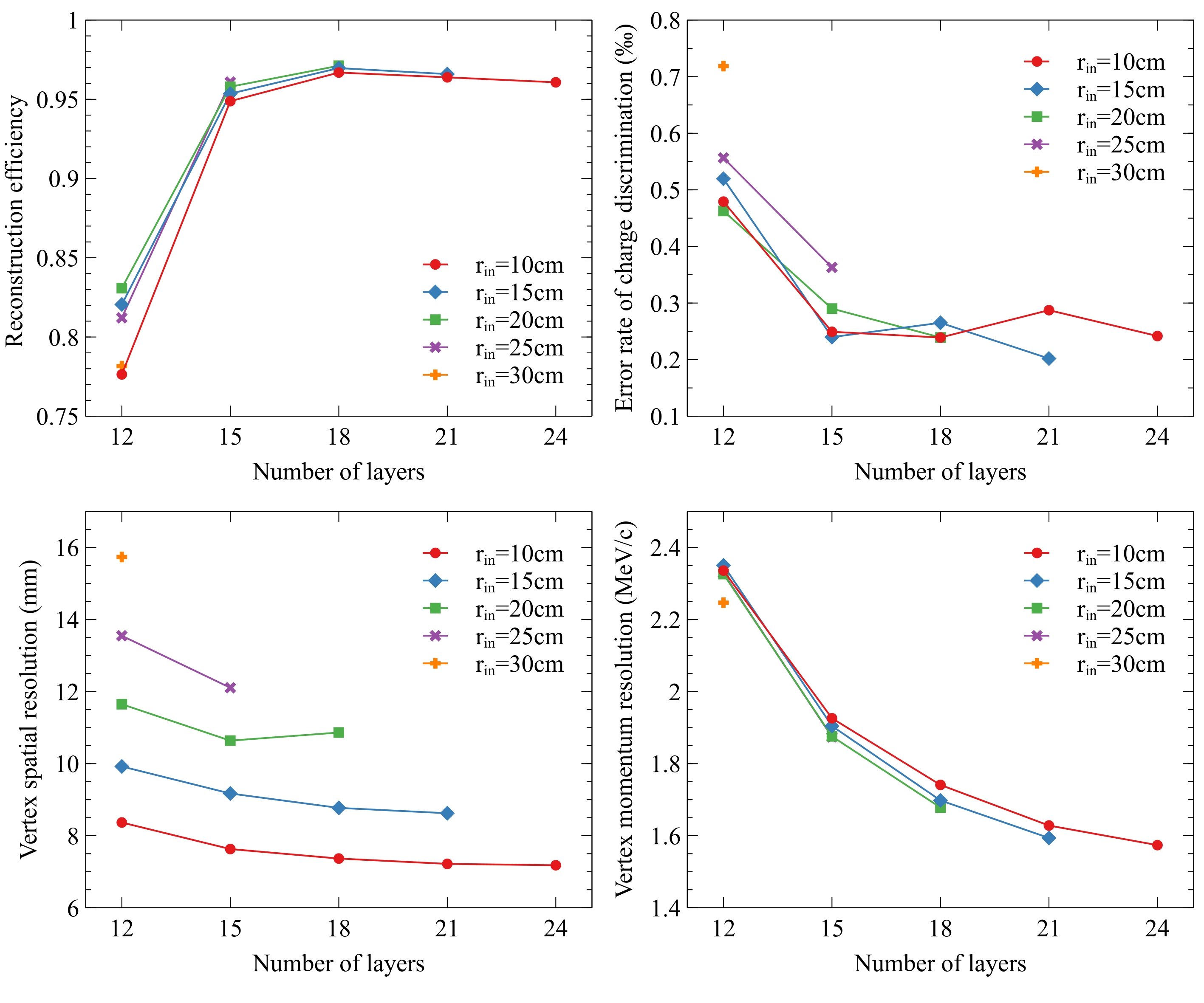}
    \caption{Tracking performance of investigated drift chamber setups. All setups share the same cell width of 16.2~mm, and solid angle coverage of 95\% (inner), 89\% (outer), thus layer counts also represents outer radius of the chamber. The top-left plot demonstrates the efficiency of fit under different chamber inner radius and layer counts. The bottom-left plot demonstrates the spatial resolution of  reconstructed decay vertex, that raises quickly with smaller inner radius, reaches the minimum of a few~mm. The vertex resolution can be further improved with the coincidence of MCP. The bottom-right plot demonstrates the reconstructed momentum resolutions. After comprehensively considering the optimal performance metrics, the practicability, and possible future upgrades, the design of 18 layers with inner radius of 15cm (setup 8 of TABLE~\ref{table:CDC parameters}) is adopted.}
    \label{fig:cdcPerformace}
\end{figure}

\begin{figure}[!t]
    \centering
    \subfigure[Exact energy spectrum.]{
        \label{fig:subfig:genEnergySpectrum}
        \includegraphics[width=8cm]{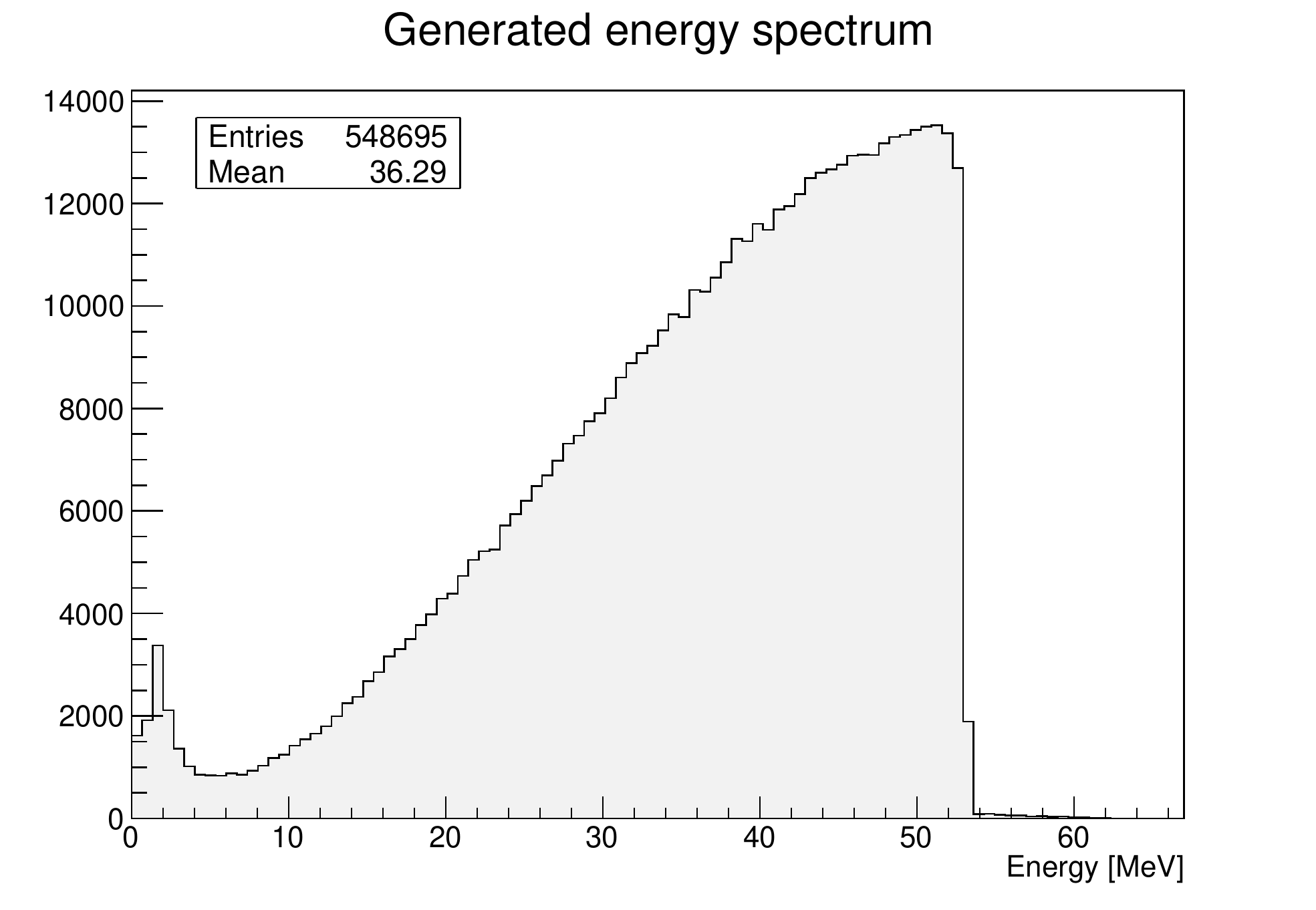}}
    \subfigure[Energy spectrum reconstructed from simulated drift chamber hits.]{
        \label{fig:subfig:recEnergySpectrum}
        \includegraphics[width=8cm]{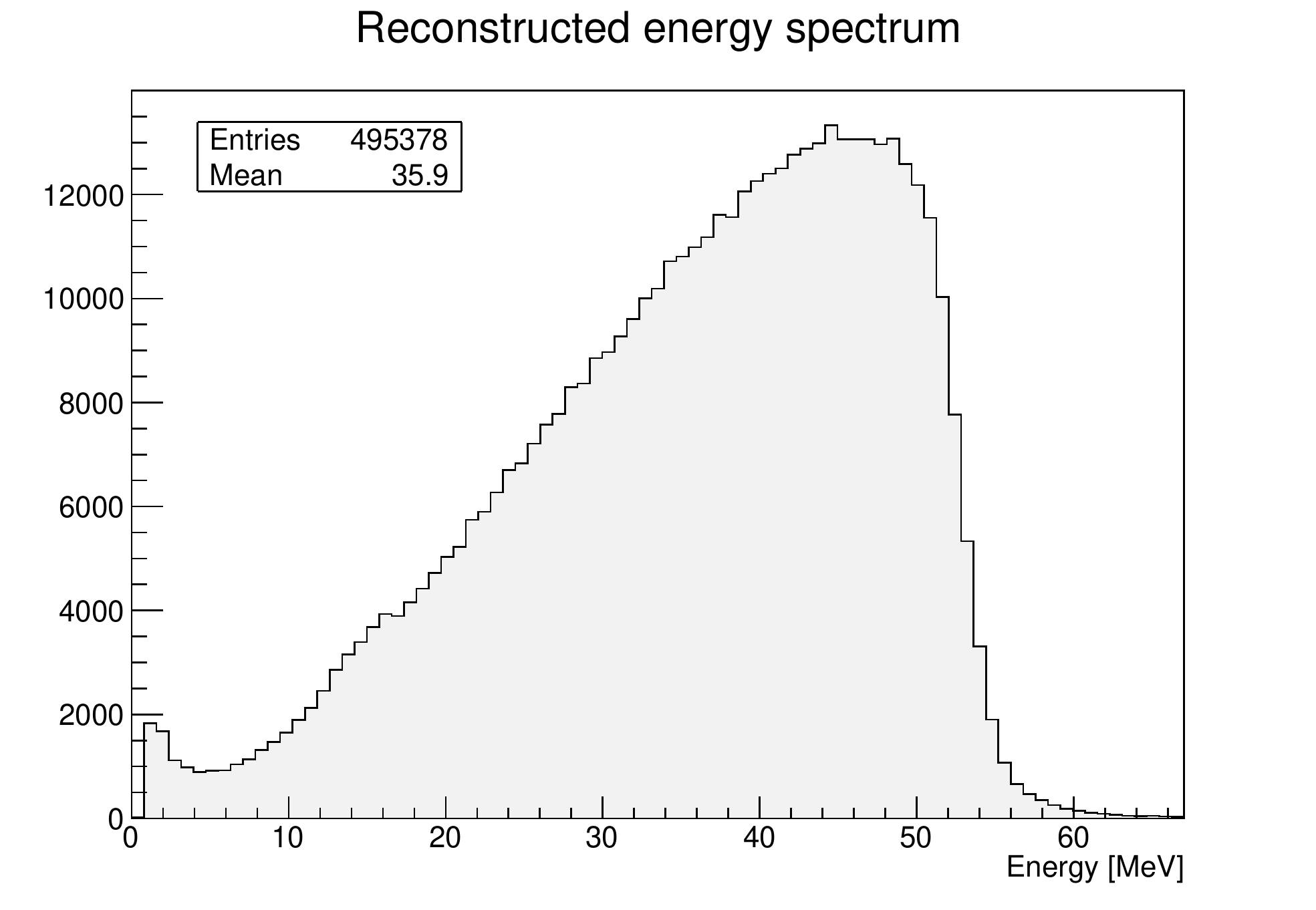}}
    \caption{Comparison between true energy spectrum and reconstructed spectrum. }
    \label{fig:energySpectrum}
\end{figure}

As for tracking, since the hodoscope and trigger system is not much investigated, the practical track finding algorithm is not yet applied. Instead, simulated hits are simply classified by event ID and track ID, and hits that scattered back from the chamber wall are cut, to represent the perfect finding result. Before fitting, hits are smeared by Gaussian noises of $\sigma_d=$0.2~mm and $\sigma_z=$3~mm for drift distance measurement and longitude measurement, respectively. The smear is according to the resolution of common drift chambers~\cite{Adamov:2018vin, BESIII:2009fln}. Then the smeared measurements are fitted by helix with least square method, and the helix parameters are converted to vertex position and momentum. Then the fitted parameters are compared with true values recorded from the previous simulation, performance metrics including spatial resolution, momentum resolution, and error rate of charge discrimination are calculated, respectively. The result are plotted in FIG.~\ref{fig:cdcPerformace} for different layer counts. For the least square fitter, tracking efficiency reaches maximum of 97.1\% at 18 layers, and inner radius of 20cm. The top-right plot demonstrates error rate of particle charge discrimination, which can be lower than 0.35\textperthousand{} for most setups that have more than 12 layers. With a comprehensive consideration of the optimal performance metric, the cost effectiveness, and possible future upgrade, the design of 18 layers with inner radius of 15cm (setup 8 of TABLE~\ref{table:CDC parameters}) will be adopted as the benchmark setup. The reconstructed energy spectrum and exact energy spectrum are plotted as FIG.~\ref{fig:energySpectrum}. As shown in FIG.~\ref{fig:cdcPerformace} and FIG.~\ref{fig:energySpectrum}, the reconstructed momentum resolution is still at above 1~MeV/$c$, which means that in principle, far from the resolution limit. Therefore, there is still large space for the improvement on tracking algorithm, which is one of the important subsequent subjects. 

\subsubsection{Microchannel plate}

So far, the simulation program for MCP has been under development. We have simplified the geometry of MCP considering it will be a heavy burden for computing to construct such a complex detector including millions of channels. In principle, the incident positron will cause initial ionization before annihilation. Then the electrons from initial ionization will be multiplied by the emission of secondary electrons. Then secondary electrons after multiplication will be read out by electronic system. As the response of each channels will be the same or at least similar with each other, it will be brilliant to study the amplitude and shape of the signal when a certain electron from initial ionization enter a channel. After the signal has been known for detail, the response of a single electron can definitely be accessed by a simple sampling so that the response of the whole MCP will be able to be simulated fast and precisely.

\begin{figure*}[!htbp]
    \centering
    \includegraphics[width=0.7\textwidth]{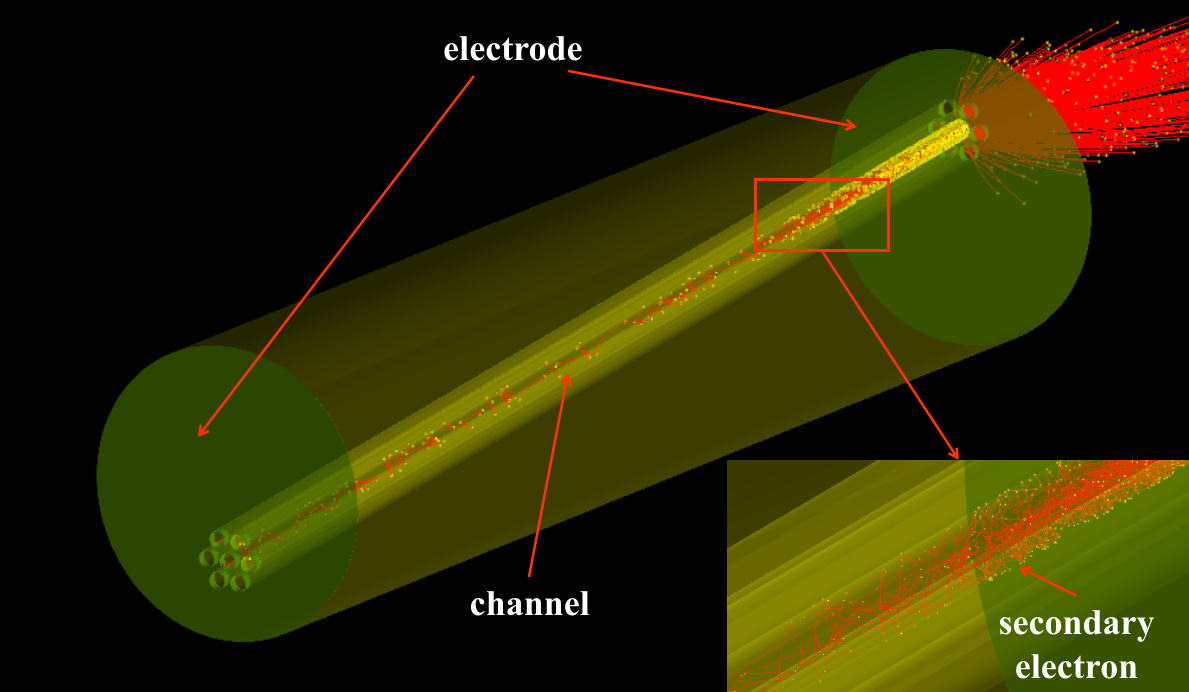}
    \caption{Simulation picture using the simple MCP model.}
    \label{fig:sim_MCP}
\end{figure*}

Therefore, a simplified model of MCP with only 7 channels has been constructed in order to get the output response of a single channel. The geometry model we used in simulation so far can be found in FIG.~\ref{fig:sim_MCP}. Fundamental parameters of MCP are provided by factory, which are shown in TABLE~\ref{tab:MCP_para}. In the picture, the main part of MCP is shown by yellow color and the green part is the electrodes. A event where an electron enters the central channel has been simulated for detail. The red lines is the tracks of secondary electrons. Limited by the rendering power of computer, only 5000 tracks are drawn in this figure. However, the amplification of secondary electrons can be clearly seen.

\begin{table}[!htbp]
    \centering
    \begin{tabular}{c|c}
        \hline
        Parameter & Value \\
        \hline
        Thickness & 0.48 mm \\
        Radius of channel & 3.0 $\rm{\mu m}$ \\
        Angle of inclination & $5.5^{\circ}$ \\
        Distance between channels & 8.0 $\rm{\mu m}$ \\
        Thickness of electrode & 0.2 $\rm{\mu m}$ \\
        Length of electrode in channels & 3.0 $\rm{\mu m}$ \\
        High voltage & 800 V \\
        \hline
    \end{tabular}
    \caption{Fundamental parameters of MCP simple model.}
    \label{tab:MCP_para}
\end{table}

\begin{figure}[!htbp]
    \centering
    \subfigure[]{ \includegraphics[width=0.45\textwidth]{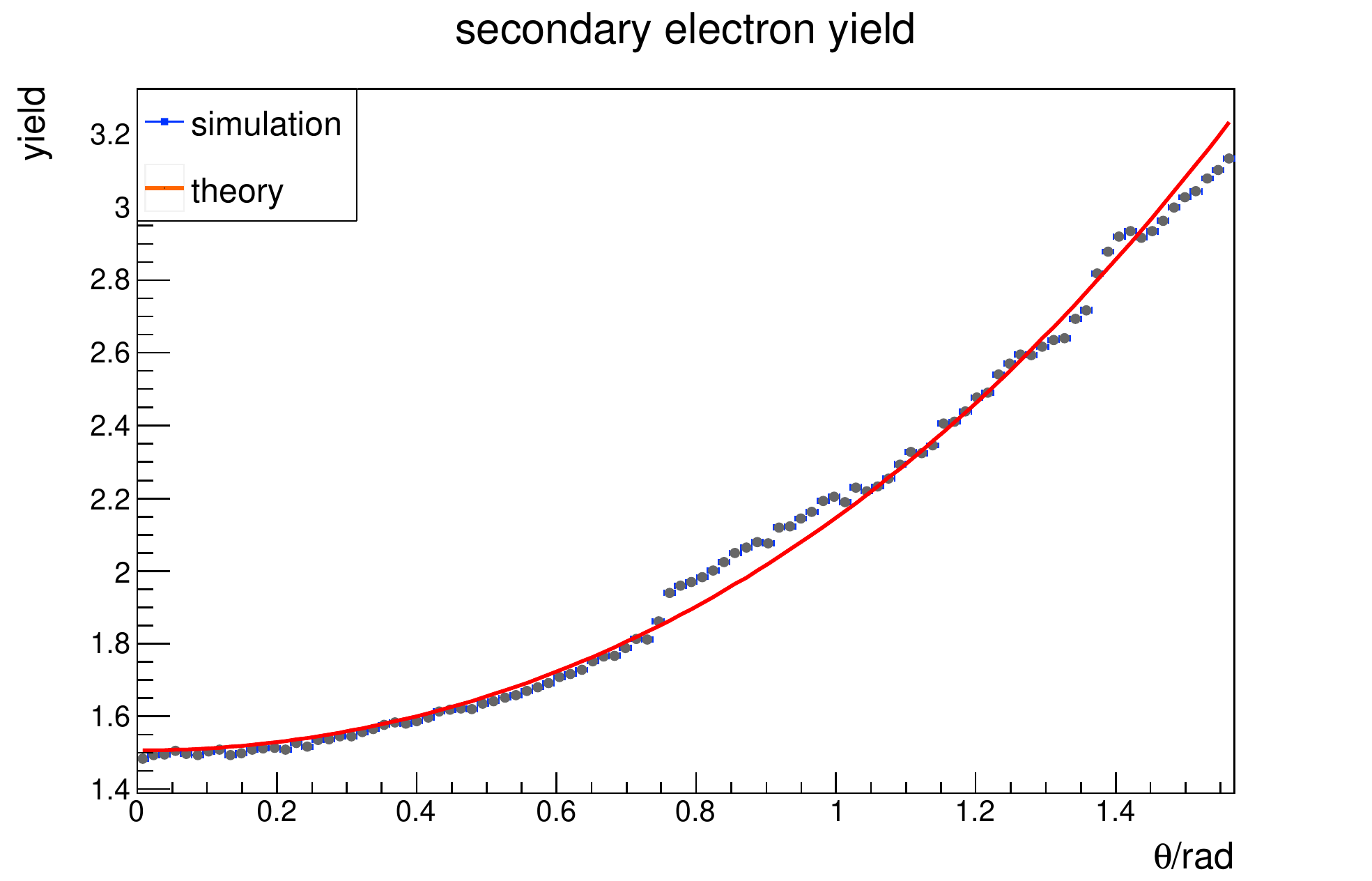} }
    \subfigure[]{ \includegraphics[width=0.45\textwidth]{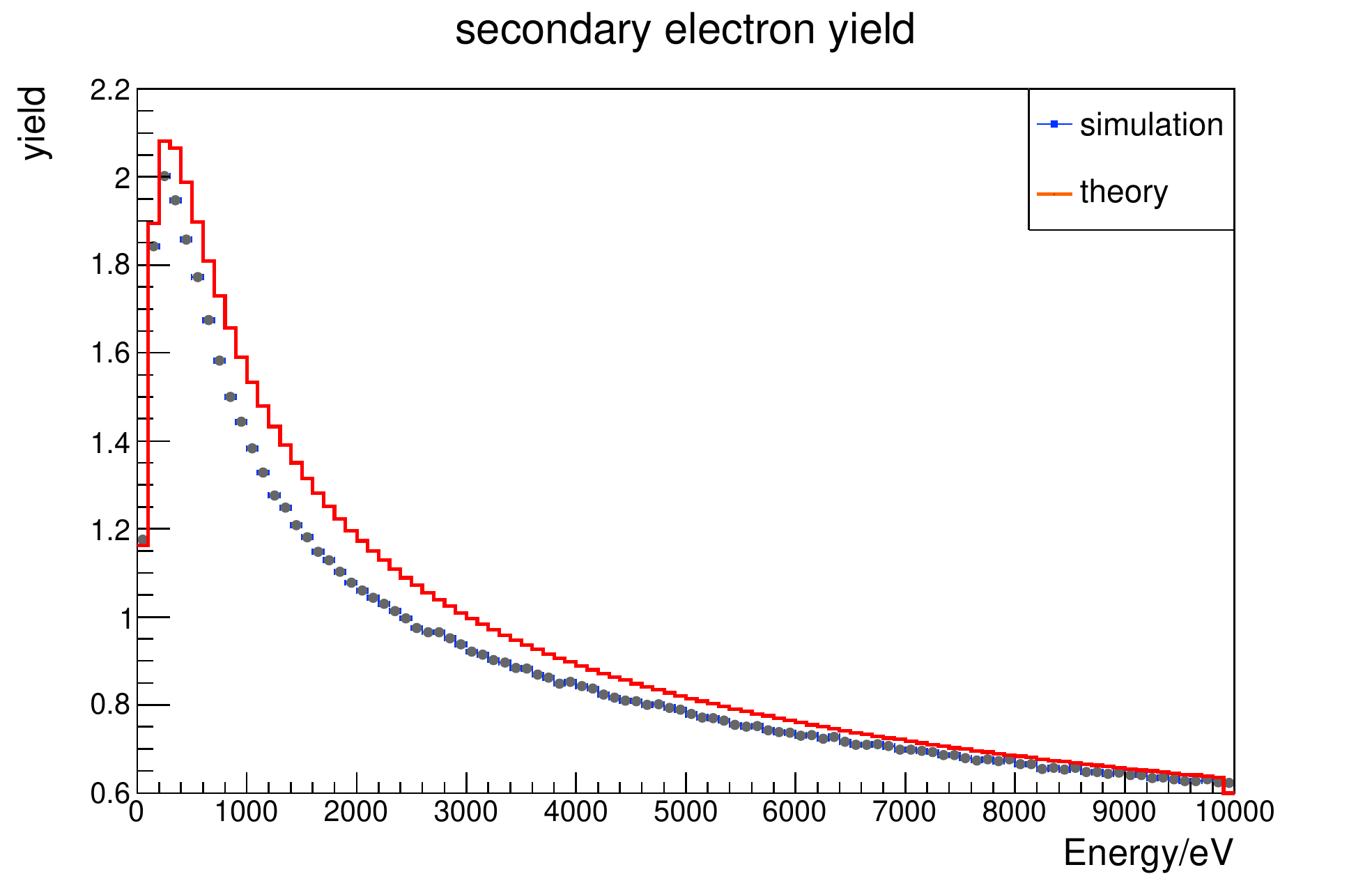} }
    \subfigure[]{ \includegraphics[width=0.45\textwidth]{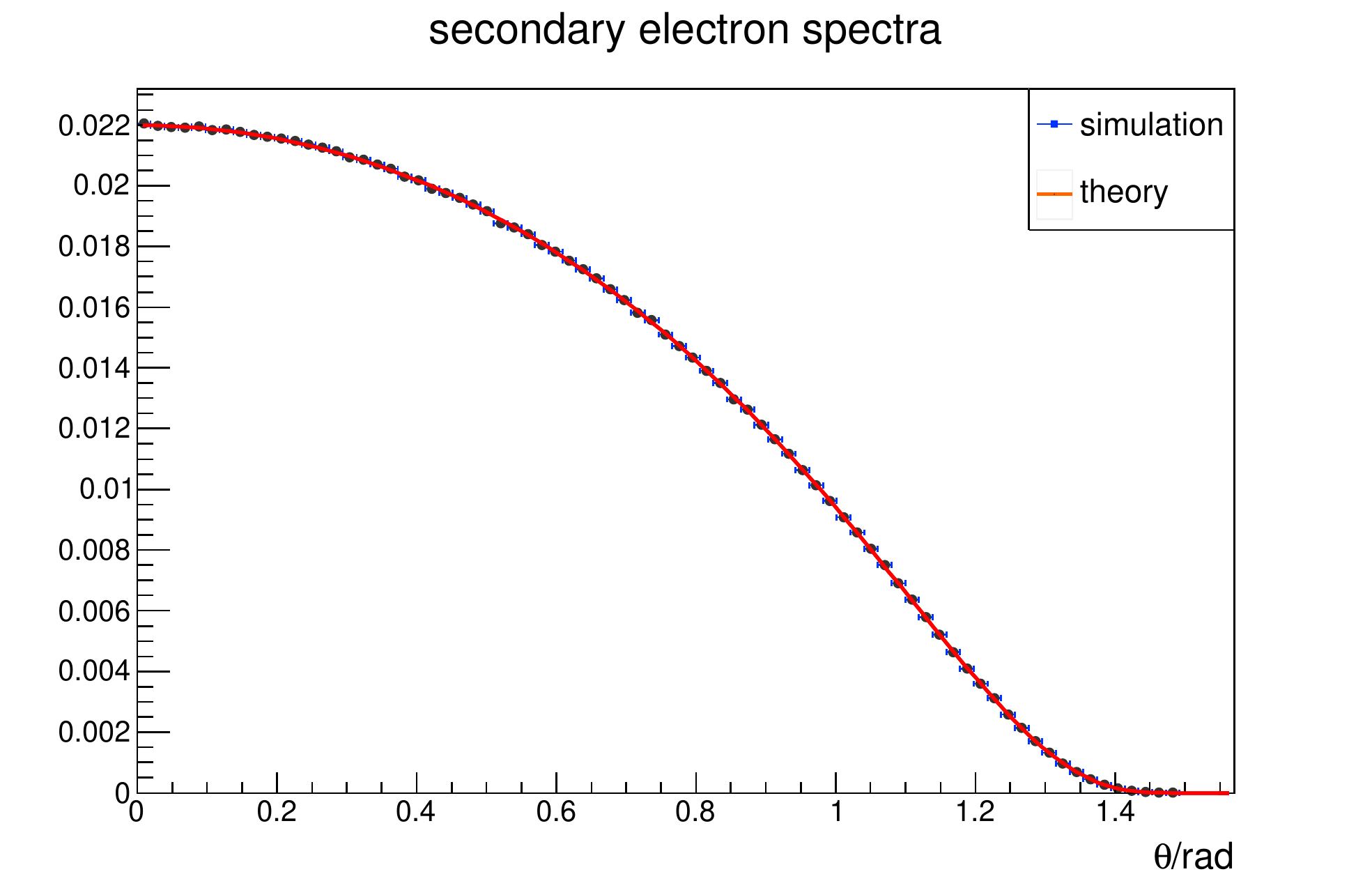} }
    \subfigure[]{ \includegraphics[width=0.45\textwidth]{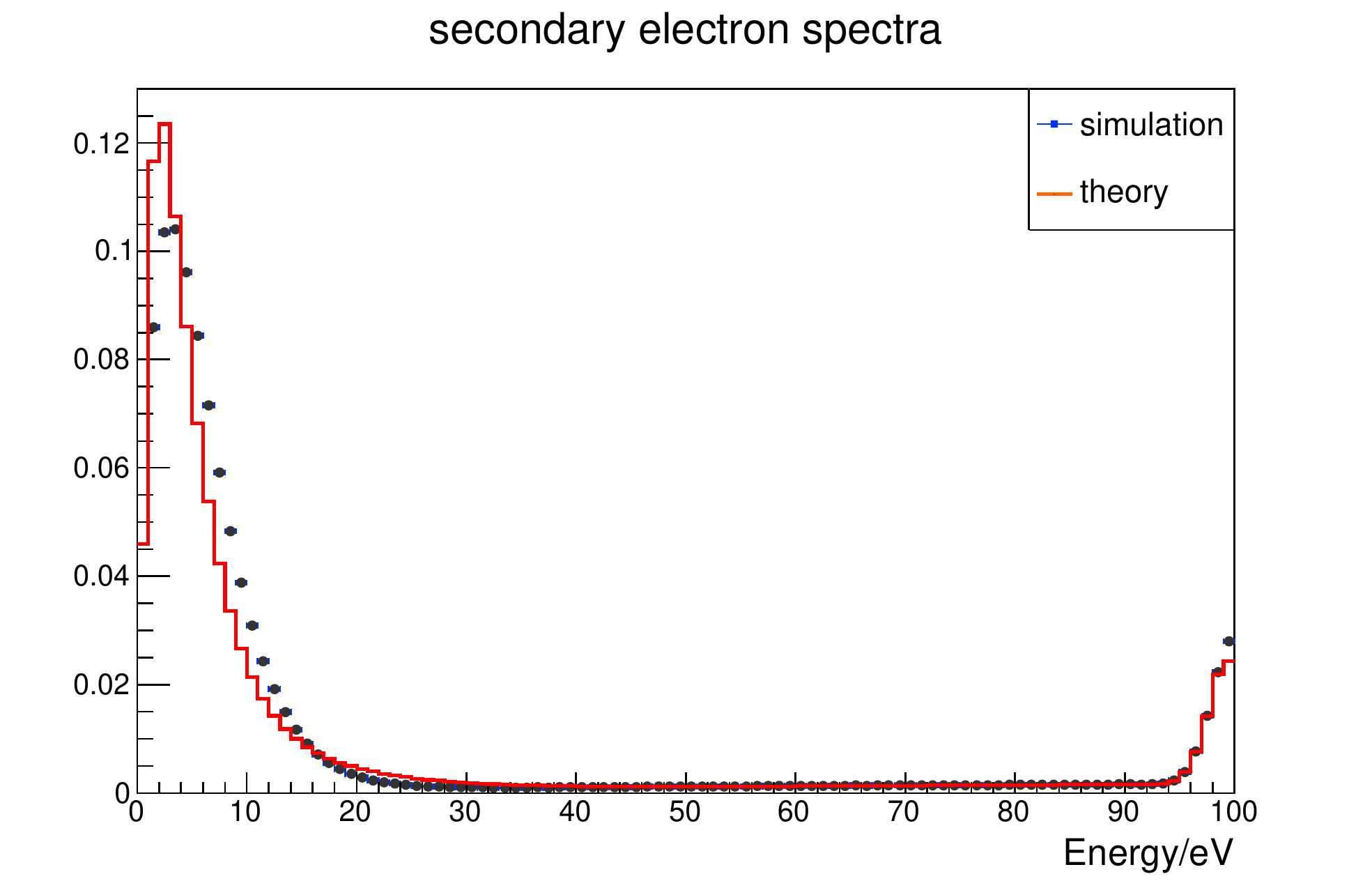} }
    \caption{Comparison between simulation results and theory model.}
    \label{fig:comparison_MCP}
\end{figure}

However, the emission of secondary electrons has not been implemented into the official code of Geant4. In order to get the precise response of MCP, we developed a generator for secondary electron according to the most well-known Furman model~\cite{Furman:2002du}, which is always used to describe the yield and angular distribution of secondary electron. In order to verify our generator, we draw several typical distributions of simulation results and theory to the same pictures, which can be found in FIG.~\ref{fig:comparison_MCP}. FIG.~\ref{fig:comparison_MCP}(a) shows the distribution of yield of secondary electrons with respect to the incident angle of initial electron, where incident angle refers to the angle between incident direction and the normal direction of geometric surface. FIG.~\ref{fig:comparison_MCP}(b) is the distribution of yield of secondary electrons with respect to the kinetic energy of incident electron. FIG.~\ref{fig:comparison_MCP}(c) and (d) are the angular and kinetic energy distribution of secondary electrons when initial electrons enter the channel with kinetic energy of 100~eV, respectively. It can be seen that simulation and theory are consistent with each other within an acceptable range although there is a little systematic difference in several distributions.

\begin{figure}[!htbp]
    \centering
    \subfigure[Time when secondary electrons arriving electronics]{ \includegraphics[width=0.45\textwidth]{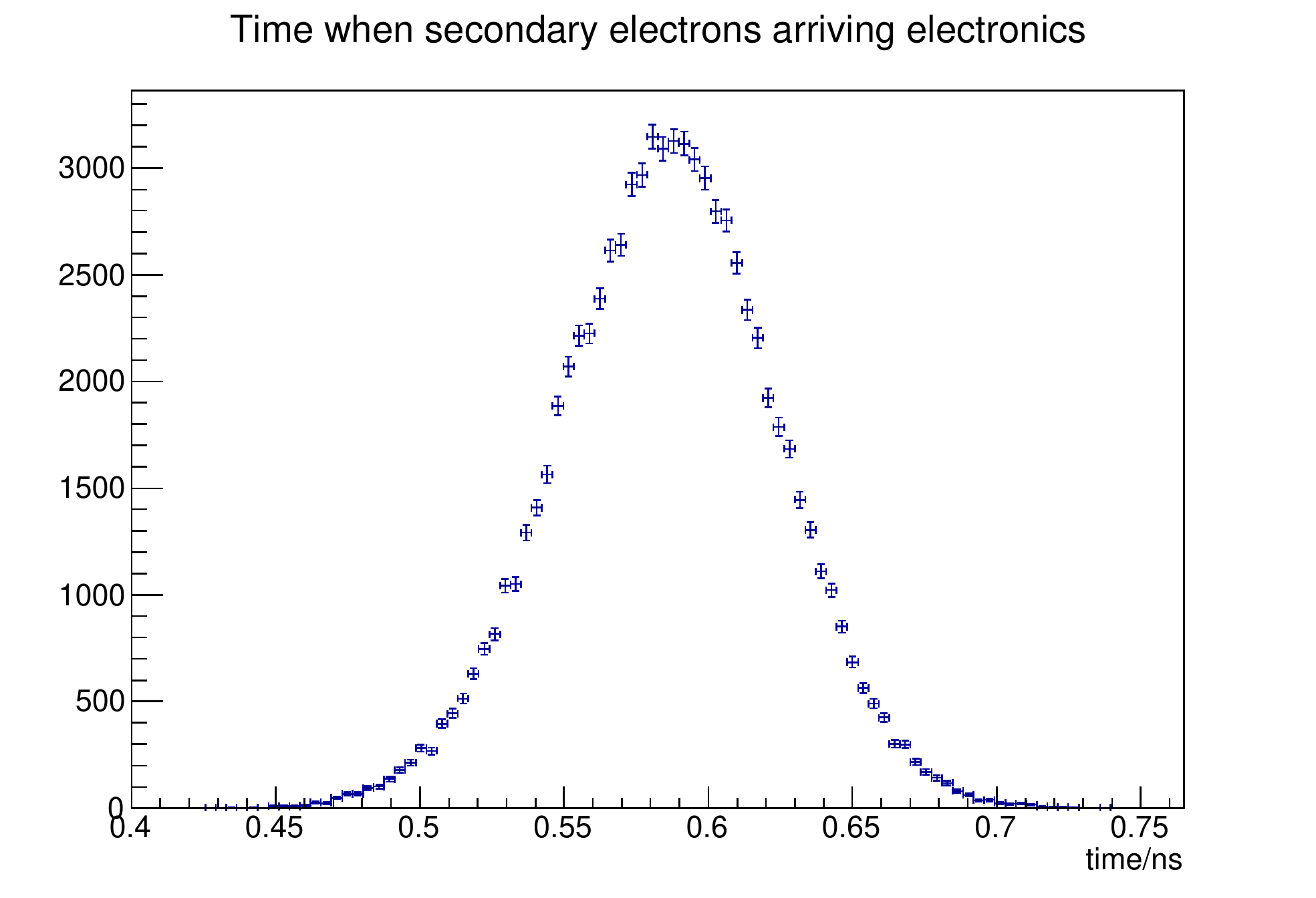} }
    \subfigure[Position when secondary electrons arriving electronics]{
    \includegraphics[width=0.45\textwidth]{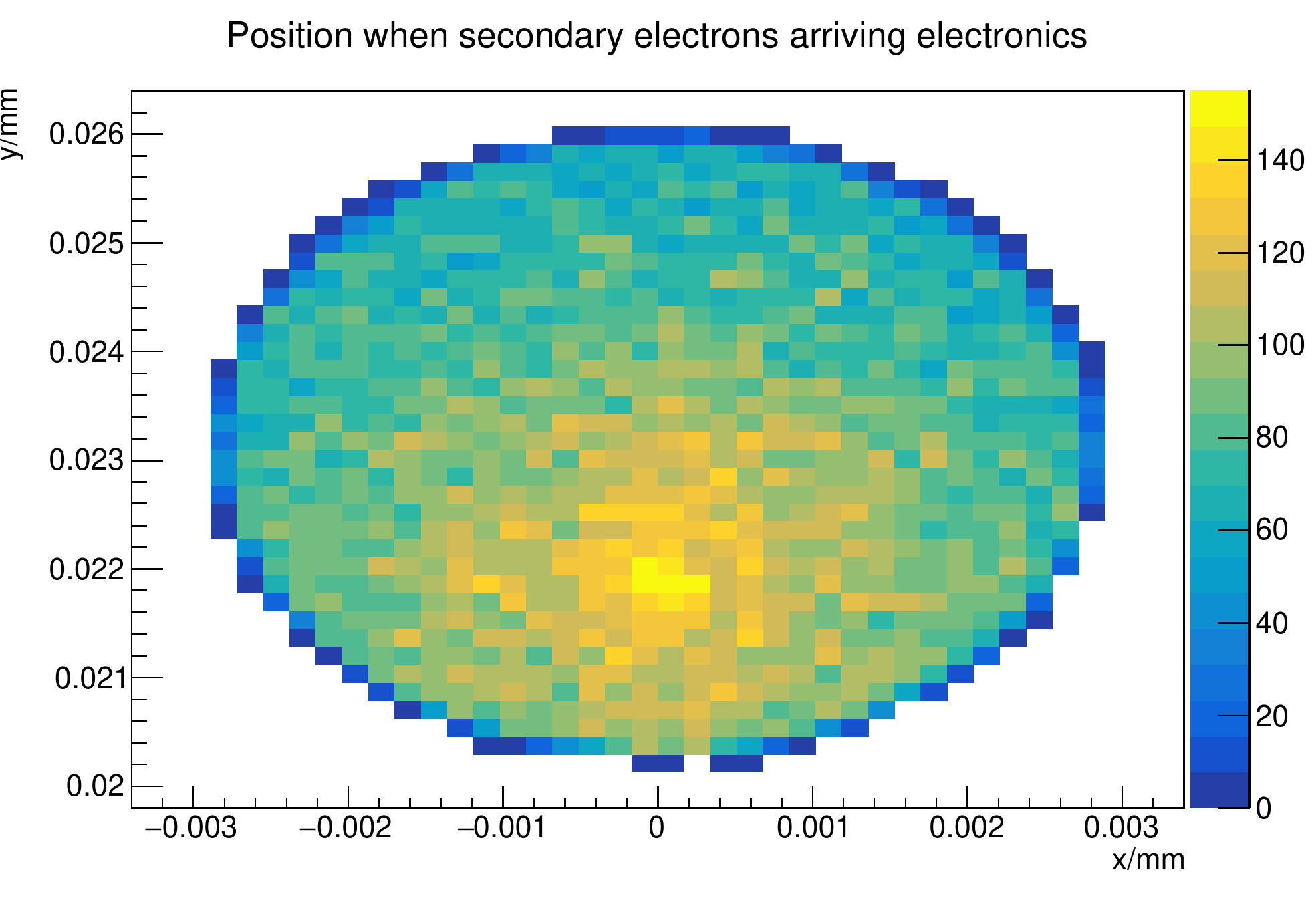} }
    \caption{Time and position distribution of secondary electrons received by electronics.}
    \label{fig:MCP_output}
\end{figure}

The output response can be got when a certain electron enter the channel of MCP after the geometry and process are defined. FIG.~\ref{fig:MCP_output} shows the output wave shape when a electron enter the channel with 100~eV kinetic energy.

There are still works to do during the design of the positron detector and a plan to improve the technical target will be brought forward in the near future.

\subsection{Discrimination of signals and backgrounds}
Two signals can be applied to identify an antimuonium which has been converted from the formed and diffused muonium in vacuum: one is the energetic electron from a $\mu^-$ decay in the magnetic spectrometer; the other is the atomic shell $e^+$. The magnetic spectrometer is the central detector component to fulfill the charge identification for final states. In addition, it is essential to reconstruct the interaction vertex. Both timing and position reconstruction have to be registered in an extremely high precision in the coincident detection techniques.

However, several backgrounds can lead to fake events. First, there might be accidental coincidence between an energetic $e^-$ produced by Bhabha scattering of $e^+$ from an uncaptured $\mu^+$ decay and a scattered $e^+$ in the $\mu^+$ beam. Cosmic ray muons might also make contributions to conincident backgrounds. Second, apart from the dominant three-body decay from $\mu^+$, we have to be confronted with the rare decay processes predicted by standard model:
\begin{equation}
 \mu^+\to e^+ + \nu_e + \bar{\nu}_\mu + e^+ + e^-\,.
\end{equation}
Based on the latest measurement in Particle Data Group~\cite{ParticleDataGroup:2020ssz}, the branch ratio is $(3.4\pm0.4)\times10^{-5}$. Both $e^-$ and $e^+$ in such a rare process show up in continuous energy spectra and can result intrinsic backgrounds. Can we carefully measure these events or subtract them in the region of interests. Or can we check the detector response in the proposed spectrometer to discriminate them from our muonium or antimuonium signals? From experimental point of view, fortunately, the measured branching ratio will become very small after a convolution of the detector energy resolution as was given in the previous study~\cite{Djilkibaev:2008jy,Flores-Tlalpa:2015vga}. Furthermore, the next-leading order calculations decrease the tree-level predictions~\cite{Pruna:2016spf, Fael:2016yle}. Whether it will contribute to coincident backgrounds will be further explored after a full consideration of physics processes in the MC simulation~\cite{Ulrich:2020frs}. Based on preliminary simulation results, we can set up the region of interests in the plane for the distance of closest area determined by the magnetic spectrometer and the time of flight by tracing the shell positrons in the linac and the MCP target back to the original interaction point. It is still under development to refine the detector system to reach the desired improvement in the physics sensitivities. 

\section{Summary}
\label{Sec:summary}
The observation of lepton flavor violation processes would indicate physics beyond the standard model. The cLFV in processes such as the $\mu^-\to e^-$ conversion, the muon decays $\mu^+\to e^+\gamma$ and $\mu^+\to e^+ e^- e^+$, and the muonium to antimuonium conversion $\mu^+ e^-\to \mu^- e^+$ may be large enough to be within the reach of future experiments. With the proposed MACE concept, it is promising to expect searches for new physics with the muonium to antimuonium conversion process. At the same time, it is important to understand the feasibility of such experiment by a scrutiny of SM rare background processes and optimizing the detector system to discriminate signals from backgrounds. We have presented progress and prospects in theoretical and experimental aspects. Together with other flavor and collider searches, MACE will shed light on the mystery of the neutrino masses.

\bibliographystyle{unsrt}
\bibliography{bibliography} 

\end{document}